\documentclass[letterpaper]{jpconf}
\usepackage{graphicx}
\begin{document}
\title{Data driven background determination for SUSY searches with ATLAS}

\author{Aleksej Koutsman on behalf of the ATLAS collaboration}

\address{ {\em Nikhef National Institute for Subatomic Physics, Science Park 105, P.O. Box 41882, NL - 1009 DB Amsterdam, Netherlands} }

\begin{abstract}
Supersymmetry (SUSY) is an attractive extension of the Standard Model possibly solving many standing issues in particle physics and cosmology. The general purpose ATLAS detector at the Large Hadron Collider (LHC) is an experiment capable of discovering or excluding TeV SUSY. However discovery can only be claimed when the Standard Model backgrounds are understood and are under control. The expectations at the LHC are that Monte Carlo simulation predictions may not be sufficient to achieve this and the backgrounds will have to be determined from data itself. In this note we will highlight some data driven methods developed to estimate backgrounds and detect a possible SUSY excess.
\end{abstract}

\section{Introduction}
In order to prevent rapid proton decay a new quantum number called R-parity is conserved in SUSY models, with two clear consequences for SUSY searches with ATLAS. The first one is the lightest supersymmetric particle is absolutely stable, giving a signature of missing energy in the detector. 
Secondly we expect SUSY events to exhibit relatively large multiplicity of high $p_\mathrm{T}$ jets, as SUSY particles with strong interaction couplings have highest production rates at the LHC and these will go through long decay chains. Furthermore it was decided to study SUSY with different search strategies based on exclusive requirement of zero, one or more leptons.
\\
\indent
The aim of the described studies was to develop data driven techniques to estimate the major Standard Model (SM) backgrounds such as W/Z bosons with associated jets, top quark pair production and jet production from QCD processes. SUSY should show up as an excess over the predicted SM events in the so-called ``signal'' region, where new physics may be present. The prediction is done by extrapolating from a ``control'' region, which should be as close as possible to the signal region, give an unbiased estimate, have sufficient statistics and small theoretical uncertainties. The first requirement has the effect of contaminating the control region with SUSY events, which results in overestimating the SM background in the signal region, but this contamination can be taken into account.
\\
\indent
 We assumed the LHC running with $\sqrt{s}=14~\mathrm{TeV/c^2}$ and a collected integrated luminosity of $1~\mathrm{fb^{-1}}$. Due to space limitations the methods described in this note are a subset of all the methods developed and the reader is referred to \cite{cscbook} for a complete description.

\section{One-lepton search mode}
The one-lepton search mode requires an isolated lepton. This will strongly suppress the QCD background that dominates over other processes by orders of magnitude and has high theoretical and instrumental uncertainties. A lepton in our studies is either an electron or a muon with $p_\mathrm{T}$ of at least $20~\mathrm{GeV/c}$. To avoid overlap with the di-lepton mode, we veto events with a second identified lepton with a $p_\mathrm{T}$ of more than $10~\mathrm{GeV/c}$. Assuming R-parity conservation we demand at least four jets with $|\eta|<2.5$ and $p_\mathrm{T}>50~\mathrm{GeV/c}$ out of which one jet must have $p_\mathrm{T}>100~\mathrm{GeV/c}$. The missing transverse energy $E^{\mathrm{miss}}_{\mathrm{T}}$ should be larger than $100~\mathrm{GeV/c}$ and above $0.2M_\mathrm{eff}$, where $M_\mathrm{eff}$ is the effective mass. Our last selection criterium is that transverse sphericity $S_\mathrm{T}$ is larger than $0.2$. All definitions of variables can be found in \cite{cscbook}.

\subsection{\bf {\em Combined fit method}}
After the selection the only non-negligible backgrounds left are W bosons with associated jets (W+jets) and top quark pairs. The latter we divide into two categories: semileptonic ($t \bar{t} \rightarrow b \bar{b} \ell \nu q \bar{q}$) and dileptonic ($t \bar{t} \rightarrow b \bar{b} \ell \nu \ell \nu$) top quark pairs, as these have different shapes and yields. Dileptonic top quark pair events end up in our one lepton sample because either the second lepton was mis-/not reconstructed or the second lepton was a tau lepton decaying hadronically, respectively constituting one third and two-thirds of all dileptonic events.

\begin{figure}[b!]
\centering\includegraphics[width=.9\linewidth]{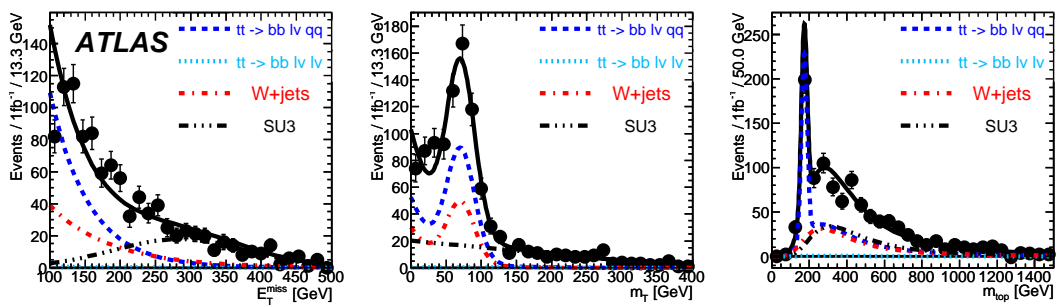}
\caption{Distribution of $E^{\mathrm{miss}}_{\mathrm{T}}$ (left), $m_\mathrm{T}$ (center) and
$m_{top}$ (right) of a 1 fb$^{-1}$ mix of $~t \bar{t}$ and W+jets SM events and 
SUSY SU3 events overlaid with projections of the
combined model fitted to the mix of
events with floating yield and shape parameters. For
each projection the contributions of the semileptonic $t \bar{t}$
contribution (dark blue), dileptonic $t \bar{t}$ contribution (light
blue), W+jets contrubution (red) and Ansatz SUSY constribution (black)
are shown.}
\label{fig:combfit-fitfixedshape}
\end{figure}

We fit the SM backgrounds in three observables: $E^{\mathrm{miss}}_{\mathrm{T}}$, $M_\mathrm{T}$ and  $m_{top}$. $M_\mathrm{T}$ is the invariant mass of $E^{\mathrm{miss}}_{\mathrm{T}}$ vector and the lepton $p_\mathrm{T}$, while $m_{top}$ is the invariant mass of the three jets with largest vector-summed $p_\mathrm{T}$.
\\ 
\indent
Taking physics features into account we construct probability density functions (p.d.f.'s) that model each of the three main backgrounds in the three observables. For example the semileptonic $t \bar{t}$ shows a clear peak in $m_{top}$ if we find the correct three quarks from top quark decay.
\\
\indent
For a broad range of SUSY parameters it was observed that the shapes of low $E^{\mathrm{miss}}_{\mathrm{T}}$ and low $M_\mathrm{T}$ distributions have little dependence with the chosen model point. Thus we can construct a model-independent Ansatz shape to describe the SUSY contamination at low energy.
\\
\indent
The combined model used in the fit is the addition of p.d.f.'s of each background sample and the SUSY Ansatz shape with a yield parameter for each separate component. The first step in the procedure is to fit the $E^{\mathrm{miss}}_{\mathrm{T}}$, $M_\mathrm{T}$ and  $m_{top}$ distributions with shapes obtained from Monte Carlo simulation. To make it more data-driven is to release in the fit as many of the shape parameters as possible. Figure \ref{fig:combfit-fitfixedshape} shows the result of the fit with floating yields and floating shape parameters to $1 ~\mathrm{fb^{-1}}$ of data.
\\
\indent
The final step is to extrapolate the yields of the SM background components to the signal region. Table \ref{tab:combfit-yieldextrap-fixed} shows the extrapolated yields to the signal region (SIG), defined by $E^{\mathrm{miss}}_{\mathrm{T}}>200~\mathrm{GeV/c}$ and $M_\mathrm{T}>150~\mathrm{GeV/c^{2}}$, while propagating all correlated parameter uncertainties with fixed and floating shapes. For comparison the same table shows extrapolated yields to the full parameter space (FULL) from the control region. The yields that we find are in good agreement with the truth values of the fitted event mix within statistical uncertainties.
	
\begin{table}[t]

\label{tab:combfit-yieldextrap-fixed}
\begin{center}
\caption{Yields from the combined fit with either fixed or floating shapes
extrapolated to the full parameter space, the truth 
yields in full parameter space, the extrapolated yields into the 
signal region and the truth yields in the signal region.}
\begin{tabular}{|l||c|c|c||c|c|c|} \hline
Component & \multicolumn{2}{|c|}{Extrap. Yield in FULL} & True & \multicolumn{2}{|c|}{Extrap. Yield in SIG} & True \\ \cline{2-3} \cline{5-6}
  & Shape Fixed & Shape Floating & FULL & Shape Fixed & Shape Floating & SIG \\
\hline
$W +{\rm jets}$         	& $ 205 \pm 45 $ & $227 \pm 68$  & 173  & $ 0.5 \pm 0.4 $ & $-1.2 \pm 2.7$  & 2 \\
1-lepton $t \bar{t}$	& $ 476 \pm 35 $ & $485 \pm 59$  & 502  & $ 0.4 \pm 0.2 $ & $-1.1 \pm 3.9$  & 0 \\
2-lepton $t \bar{t}$	& $ 62 \pm 38 $ & $17 \pm 54$  & 70  &  $ 4.5 \pm 2.9 $ & $4.7 \pm 7.9$ & 5 \\
SUSY SU3         	& $ 273 \pm 33 $ & $287 \pm 38$ & 271  & $ 92.7 \pm 2.8 $ & $95.6 \pm 4.0$  & 91 \\
\hline
\end{tabular}
\end{center}
\end{table}

\subsection{\bf {\em HT2 method}}
If we add an extra cut of $M_\mathrm{T}>100~\mathrm{GeV/c^{2}}$ to our event selection the only significant background we are left with is dileptonic $t \bar{t}$. The HT2 method estimates this background by using two near independent variables HT2 and $E^{\mathrm{miss}}_{\mathrm{T}}$ significance defined as:

\begin{center}
\begin{equation}
 \mathrm{HT2} \equiv \sum_{i=2}^4 p_\mathrm{T}^{\mathrm{jet}\, i} + p_\mathrm{T}^{\mathrm{lepton}},\;\;\;\;\;\; E^{miss}_{\mathrm{T}} \mathrm{significance} \equiv E^{miss}_{\mathrm{T}}/[\mathrm{0.49} \cdot \sqrt{\sum E_\mathrm{T}}]
\label{eq:ht2}
\end{equation}
\end{center}

\begin{figure}[b]
  \begin{minipage}{0.5\linewidth}
  \begin{center}
    \includegraphics[width=0.85\linewidth]{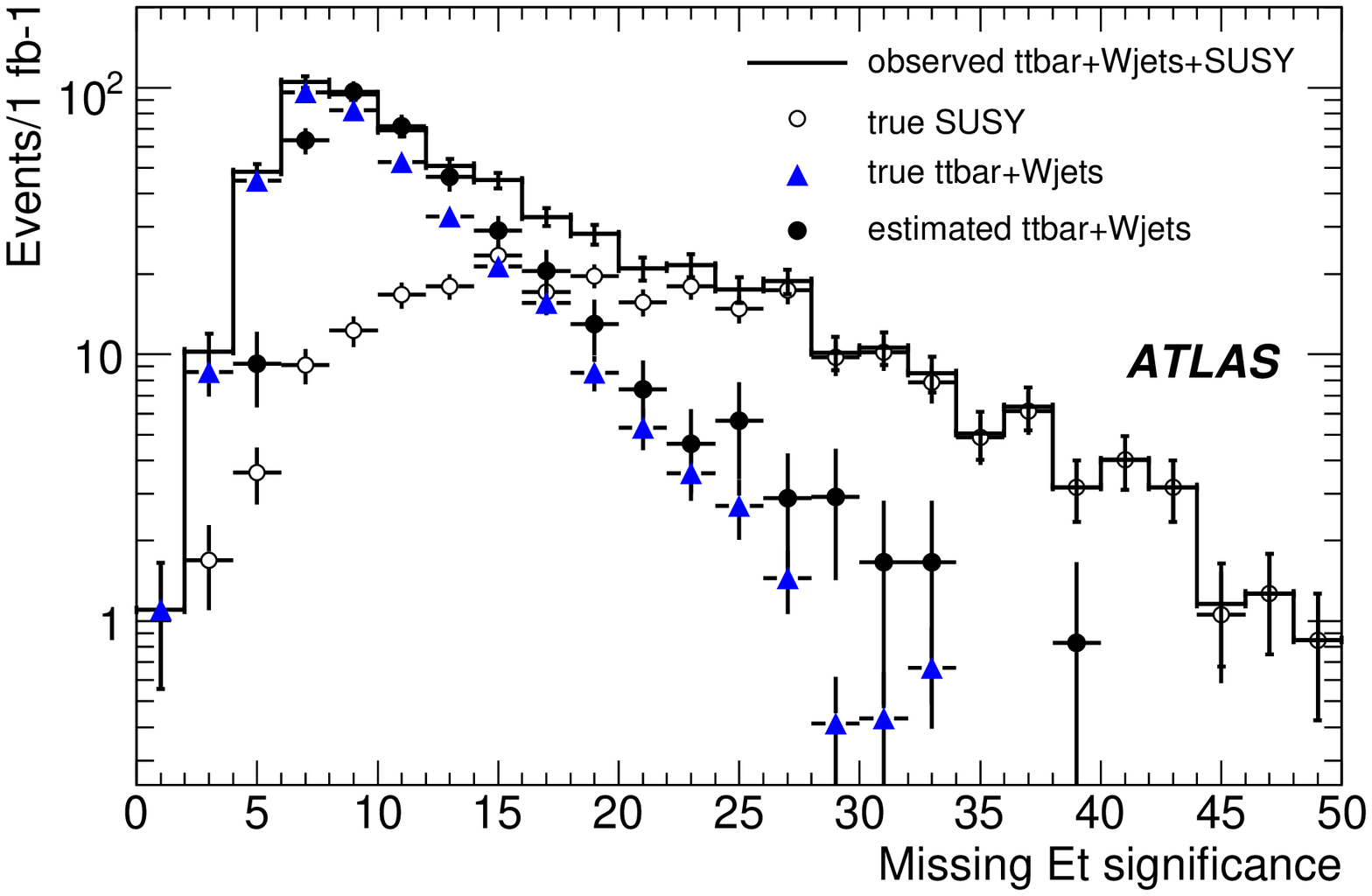}
  \end{center}
  \end{minipage}\hfill
  \begin{minipage}{0.5\linewidth}
  \begin{center}
    \includegraphics[width=0.85\linewidth]{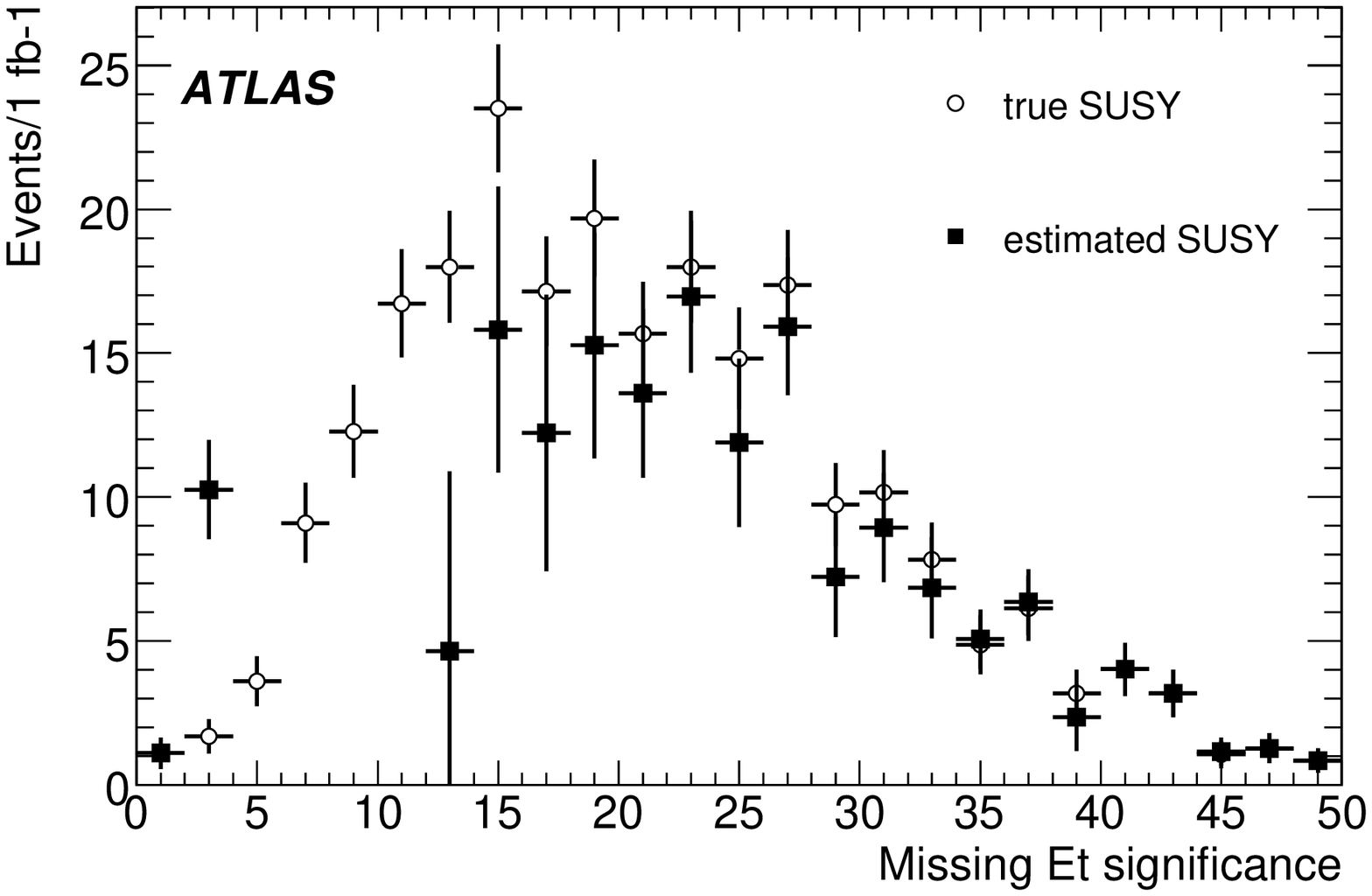}
  \end{center}
  \end{minipage}\hfill  
  \caption{{\bf Left:} Histogram is observed $E^{\mathrm{miss}}_{\mathrm{T}}$ significance
    distribution, open circles are true SUSY signal, blue triangles are true SM
    background, black filled circles is the estimated background.   
	{\bf Right:} Open circles is true SUSY signal as a
	    function of $E^{\mathrm{miss}}_{\mathrm{T}}$ significance, black is estimated SUSY excess obtained by subtracting
 	   the estimated background from the observed $E^{\mathrm{miss}}_{\mathrm{T}}$ significance distribution.}
  \label{fig:metsig_SUSY1TeV}
\end{figure}

The leading jet $p_\mathrm{T}$ was excluded from HT2 in order to reduce the correlation with $E^{\mathrm{miss}}_{\mathrm{T}}$. The correlation between the highest-$p_\mathrm{T}$ jet and $E^{\mathrm{miss}}_{\mathrm{T}}$ is a consequence of kinematics, as the rest of the event recoils from this jet. As the $E^{\mathrm{miss}}_{\mathrm{T}}$ resolution depends on ${\sum E_\mathrm{T}}$ that is clearly related to HT2,  $E^{\mathrm{miss}}_{\mathrm{T}}$ significance was used instead of $E^{\mathrm{miss}}_{\mathrm{T}}$ to further remove any correlation.
\\
\indent
A control sample is defined by HT2$<300~\mathrm{GeV/c}$ from which the shape of $E^{\mathrm{miss}}_{\mathrm{T}}$ significance is taken. The assumption is that this shape is independent of HT2. The normalization of this shape is obtained by comparing the event counts of the control sample to the signal sample at $8<E^{miss}_{\mathrm{T}} \mathrm{significance} <14$. This low $E^{\mathrm{miss}}_{\mathrm{T}}$ significance region is equivalent to a low $E^{\mathrm{miss}}_{\mathrm{T}}$ region almost unpopulated by SUSY.
\\
\indent
The background distribution and the excess of SUSY signal in high $E^{\mathrm{miss}}_{\mathrm{T}}$ significance is shown in Figure \ref{fig:metsig_SUSY1TeV}(left). The background is somewhat overestimated due to SUSY contamination in the control sample. However the right histogram of Figure \ref{fig:metsig_SUSY1TeV} shows that if we would cut higher on $E^{\mathrm{miss}}_{\mathrm{T}}$ significance, the SUSY signal would still be clearly seen over the estimated background.

\section{No-lepton search mode}
As the name suggests we veto all events with an identified muon or electron of $p_\mathrm{T}>20~\mathrm{GeV/c}$. For the rest the no-lepton mode event selection is equivalent to one-lepton mode concerning jets, $E^{\mathrm{miss}}_{\mathrm{T}}$, $S_\mathrm{T}$ and $M_\mathrm{eff}$. To pass the $E^{\mathrm{miss}}_{\mathrm{T}} > 100~\mathrm{GeV/c}$ requirement QCD events must contain a poorly reconstructed jet which can be caused by dead material, jet punch-through, pile-up of machine backgrounds and others. $E^{\mathrm{miss}}_{\mathrm{T}}$ will then point to or away from this poorly reconstructed jet. To reject these QCD events the value of the difference in azimuthal angle between the $E^{\mathrm{miss}}_{\mathrm{T}}$ vector and each of the three highest-$p_\mathrm{T}$ jets is required to be larger than $0.2$. 



\subsection{\bf {\em Replace method}}
The $Z \rightarrow \nu \bar{\nu}$ with associated jets process is one of the main backgrounds in the no-lepton search. To estimate its shapes and expected number of events $Z \rightarrow \ell^+ \ell^-$ events are selected by requiring: two opposite charged leptons, invariant mass of the two leptons within $10~\mathrm{GeV/c^{2}}$ of the Z-mass and $E^{\mathrm{miss}}_{\mathrm{T}}<30~\mathrm{GeV/c}$. Then the charged leptons are replaced by neutrinos, variables are recalculated and events go through the no-lepton selection procedure again.
\\
\indent
Four corrections must be applied to get a correct estimate, as summarized by this formula:

\begin{equation}
N_{Z \rightarrow \nu \bar{\nu}}(E^{miss}_{\mathrm{T}}) = 
\frac{N_{Z \rightarrow \ell^+ \ell^-}(p_{\mathrm{T}(\ell^+ \ell^-)})}
	{\mathrm{eff}(\eta_{\ell^+},p_{\mathrm{T},\ell^+}) \cdot \mathrm{eff}(\eta_{\ell^-}, p_{\mathrm{T},\ell^-} ) }
\times c_{\rm Kin}(p_\mathrm{T}(Z)) \times c_{\rm Fidu}(p_\mathrm{T}(Z)) 
\times \frac{{\rm Br}(Z \rightarrow \nu \bar{\nu})}{{\rm Br}(Z \rightarrow \ell^+ \ell^-)},
\end{equation}

where $N_{Z \rightarrow \nu \bar{\nu}}$ is the corrected number of events per bin of $E^{\mathrm{miss}}_{\mathrm{T}}$,
$N_{Z \rightarrow \ell^+ \ell^-}$ is the raw number of control sample events as a function of $p_\mathrm{T}(Z)$,  
$c_{\rm Kin}$ is the kinematic correction due to extra selection criteria,
$c_{\rm Fidu}$ is the fiducial correction since we cannot detect leptons beyond $|\eta|<2.5$, 
$\mathrm{eff}(\eta_{\ell^+},p_{\mathrm{T},\ell^+})$ and $\mathrm{eff}(\eta_{\ell^-}, p_{\mathrm{T},\ell^-})$ are corrections for lepton reconstruction efficiency as a function of $\eta$ and $p_\mathrm{T}$
and finally we correct for the difference in branching ratios of the Z boson. 
\\
\indent
This method does not suffer from SUSY contamination as the stringent control sample selection cuts make it practically free of SUSY. The described procedure estimates the $Z \rightarrow \nu \bar{\nu}$ background correctly but its precision is limited by statistics in the control sample.

\section{Conclusion}
What we have shown in this note are a number of methods to estimate the top quark pair, W+jets, Z+jets and QCD backgrounds from data in regions where we expect to find a SUSY excess. For an integrated luminosity of $1~\mathrm{fb^{-1}}$ many complementary methods have been developed and are shown to reliably assess the SM backgrounds as well as the possible SUSY excess. 
Ongoing work focussed on the expected early LHC running scenario with a integrated luminosity of $200~\mathrm{pb^{-1}}$ at collision energies close to $10~\mathrm{TeV/c^{2}}$ shows promising results.

\end{document}